\documentclass[preprint,5p,authoryear]{elsarticle}
\usepackage{amssymb}
\usepackage{breakcites}
\usepackage{tabularx}
\usepackage{makecell}
\usepackage{rotating}
\usepackage{threeparttable}
\usepackage{booktabs}
\usepackage{array}
\usepackage{multirow}
\usepackage{algorithm}
\usepackage{algorithmic}

\usepackage{graphicx}
\usepackage{caption}
\usepackage{subcaption}

\usepackage{verbatim}
\usepackage{lineno,hyperref}
\modulolinenumbers[5]
\usepackage{amsmath,graphicx}
\usepackage{fixltx2e}
\usepackage{amsmath,bm}
\usepackage{diagbox}
\usepackage{color}
\usepackage[varg]{txfonts}
\usepackage{siunitx}
\usepackage{ulem}
\PassOptionsToPackage{numbers,sort&compress}{natbib} 
\renewcommand\cite[1]{\citeauthor{#1}, \citeyear{#1}}
\usepackage{arydshln} 

\journal{Expert Systems with Applications}
\biboptions{authoryear}

\begin{document}
\begin{frontmatter}

\title{Perturbation Self-Supervised Representations for Cross-Lingual Emotion TTS: Stage-Wise Modeling of Emotion and Speaker}
\author{Cheng Gong$^{1,2}$, Chunyu Qiang$^{1}$, Tianrui Wang$^{1}$,
Yu Jiang$^{1}$, Yuheng Lu$^{1}$, Ruihao Jing$^{2}$, \\ Xiaoxiao Miao$^{3}$, Xiaolei Zhang$^{2,4}$, Longbiao Wang$^{1,*}$, Jianwu Dang$^{5}$ }

\address{$^1$Tianjin Key Laboratory of Cognitive Computing and Application, College of Intelligence and Computing, Tianjin University, Tianjin, China\\ $^2$ Institute of Artificial Intelligence (TeleAI), China Telecom, China \\ $^3$ Duke Kunshan University, China \\ 
$^4$ Northwestern Polytechnical University \\
$^5$ Shenzhen Institute of Advanced Technology, Chinese Academy of Sciences, Guangdong, China}
\cortext[cor1]{Corresponding author}
\ead{gongchengcheng@tju.edu.cn; qiangchunyu@tju.edu.cn; wangtianrui@tju.edu.cn; jiang\_y@tju.edu.cn; luyuheng2024@tju.edu.cn; ruihaojing@mail.nwpu.edu.cn; xiaoxiao.miao@dukekunshan.edu.cn; xiaolei.zhang@nwpu.edu.cn; longbiao\_wang@tju.edu.cn; jdang@jaist.ac.jp}

\begin{abstract}
Cross-lingual emotional text-to-speech (TTS) aims to produce speech in one language that captures the emotion of a speaker from another language while maintaining the target voice's timbre. This process of cross-lingual emotional speech synthesis presents a complex challenge, necessitating flexible control over emotion, timbre, and language.
However, emotion and timbre are highly entangled in speech signals, making fine-grained control challenging. To address this issue, we propose EMM-TTS, a novel two-stage cross-lingual emotional speech synthesis framework based on perturbed self-supervised learning (SSL) representations. In the first stage, the model explicitly and implicitly encodes prosodic cues to capture emotional expressiveness, while the second stage restores the timbre from perturbed SSL representations. We further investigate the effect of different speaker perturbation strategies—formant shifting and speaker anonymization—on the disentanglement of emotion and timbre. To strengthen speaker preservation and expressive control, we introduce Speaker Consistency Loss (SCL) and Speaker–Emotion Adaptive Layer Normalization (SEALN) modules. Additionally, we find that incorporating explicit acoustic features (e.g., F0, energy, and duration) alongside pretrained latent features improves voice cloning performance. Comprehensive multi-metric evaluations, including both subjective and objective measures, demonstrate that EMM-TTS achieves superior naturalness, emotion transferability, and timbre consistency across languages.
\end{abstract}
 

\end{frontmatter}

\section{Introduction}
Speech synthesis is a key component of the
human–computer interface that is considered essential to
responding and plays a vital role in enabling machines to generate human-like responses.
The goals of speech synthesis can be hierarchically categorized, from easier to more challenging, into three levels: intelligibility, naturalness, and expressiveness.
Speech synthesis has made significant progress in intelligibility and naturalness, mainly due to advances in deep learning and neural networks \citep {ren2021fastspeech,ju2024naturalspeech}. 
Today, we can generate speech that is often indistinguishable from human speech. 
While significant progress has been made in intelligibility and naturalness, achieving expressive and emotionally rich speech remains challenging.
And a challenging research problem persists:
cross-lingual emotion TTS \citep{10244091,guo24b_interspeech} refers
to the task of a speaker of one language to mimic the emotion of a speaker from another language while speaking a different language.

\begin{figure*}
\centerline{\includegraphics[width=1.9\columnwidth]{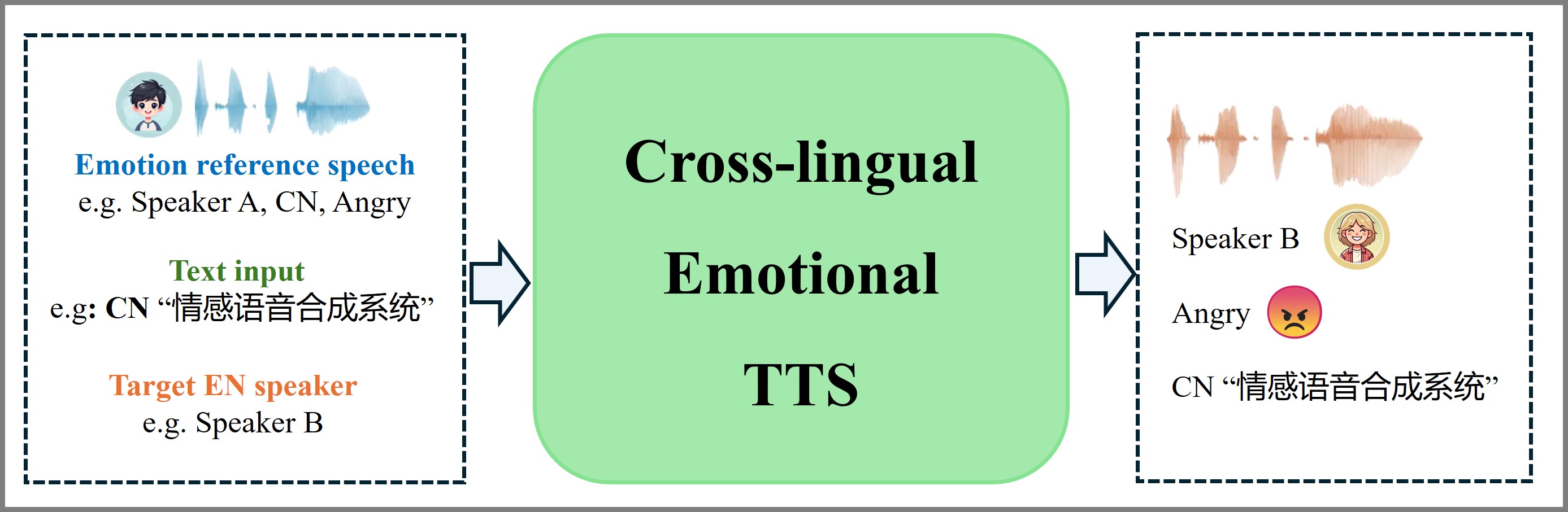}}
\caption{The problem definition of cross-lingual emotion speech synthesis.}
\label{fig:1}
\end{figure*}

Cross-lingual synthesis poses a more complex challenge in multilingual speech synthesis, as it requires transferring a speaker's voice characteristics across languages.
Despite significant efforts in cross-lingual TTS \citep{pmlr-v162-casanova22a,casanova24_interspeech} research, there remains a noticeable gap in the naturalness of generated speech compared to native speakers. 
This issue primarily arises from two factors: the lack of data resources and variations in text representations across languages.
The most straightforward approach to cross-lingual synthesis is to train the model on bilingual speech data \citep {CAI2023101427}, where the same speaker provides utterances in multiple languages.
Regrettably, collecting such bilingual data is costly, and no large-scale bilingual speech datasets are available.
Along with speech data, the lack of text resources is a major obstacle in multilingual speech synthesis.
Conventional speech processing systems that are based on phonetics require pronunciation dictionaries \citep{8682674}. These dictionaries map phonetic units to their corresponding words.
Creating such resources requires expert knowledge for each language. Despite the significant human effort involved, many languages still lack sufficient linguistic resources to develop these dictionaries.

Fortunately, the rise of self-supervised representations \citep{NEURIPS2020_92d1e1eb,9585401,9814838} has reduced the model's dependence on labeled data.
Multilingual SSL speech or text representations \citep{thenguyen23_interspeech,conneau2020unsupervised} can learn to extract linguistic, paralinguistic, and non-linguistic information from vast amounts of unlabeled data. 
Recently, they have been widely used in cross-lingual TTS to address the above issues and enhance the quality of cross-lingual TTS \citep{10669054,10.24963/ijcai.2023/575}.
Among these, ZMM-TTS \citep{10669054} integrates text-based and speech-based self-supervised learning models for multilingual speech synthesis, enabling zero-shot generation under limited data conditions. 

Over the past year, large-scale speech synthesis systems have emerged \citep{10842513,chen2024f5,du2024cosyvoice,anastassiou2024seed}, leveraging codec models and language models to significantly enhance the capabilities of voice cloning, alongside models based on self-supervised representations. 
While these models showcase impressive performance in multilingual and emotional synthesis, their focus on voice cloning and zero-shot capabilities often comes at the expense of flexible control over emotion and timbre.
Moreover, the entanglement of speaker timbre and emotion in speech may result in speaker timbre leakage during cross-speaker emotion transfer \citep{9747987}.
A common strategy for decoupling involves adversarial learning and constraints on classification losses, as demonstrated in previous research \citep{9693186,10244091}.
These methods utilize classification loss or gradient reversal to learn representations that isolate emotion or speaker information.
However, adversarial learning would introduce instability and degrade the quality of the synthesized speech. Furthermore, constraints on emotion classification may limit the emotional diversity of synthesized speech. Another straightforward decoupling approach involves speaker perturbation, which alters speaker-specific acoustic properties, such as formants, in speech \citep{10423864,9874835}. 
This perturbation method may degrade speech quality. 
Furthermore, the effects of recent speaker perturbation methods, such as speaker anonymization \citep{tomashenko2024voiceprivacy,10244064}, on speech synthesis, especially for SSL-based synthesis models, remain an underexplored area.

Motivated by the analysis above, this paper extends the previous SSL-based ZMM-TTS \citep{10669054} model by incorporating emotional speech synthesis capabilities and proposes an emotional multilingual multispeaker TTS system (EMM-TTS). The following are the major contributions of this work: 
\begin{itemize}
    \item To achieve effective decoupling of speaker and emotion, we propose a two-stage modeling approach: The first stage leverages explicit and implicit prosodic information to model emotions. In contrast, the second stage focuses on restoring the target timbre. 
    \item Additionally, we explore the effects of two different speaker perturbation methods—formant shift and speaker anonymization—on the quality of synthesized audio.
    \item To further improve speech similarity during the speech generation process, we propose a Speaker-Emotion Adaptive Layer Normalization (SEALN) and introduce a Speaker Consistency Loss (SCL).
\end{itemize} 
By comparing our proposed EMM-TTS model with the baseline, we demonstrated its effectiveness.
Audio samples can be found on our demo page.\footnote{\url{https://gongchenghhu.github.io/EMMTTS-demo/}}.

The structure of this paper is as follows: Section 2 outlines the problem we aim to address and provides a detailed explanation of our proposed method.
Section 3 details the experimental setup. Section 4 reports the experimental results. Section 5 includes the analysis and discussion.
The final section discusses related topics and summarizes the paper's contributions.
\section{Propose method}
This section will introduce the proposed EMM-TTS framework. 
We begin with fundamental knowledge about cross-lingual emotion speech synthesis, and then present the two-stage structure.

\begin{figure*}
\centerline{\includegraphics[width=1.85\columnwidth]{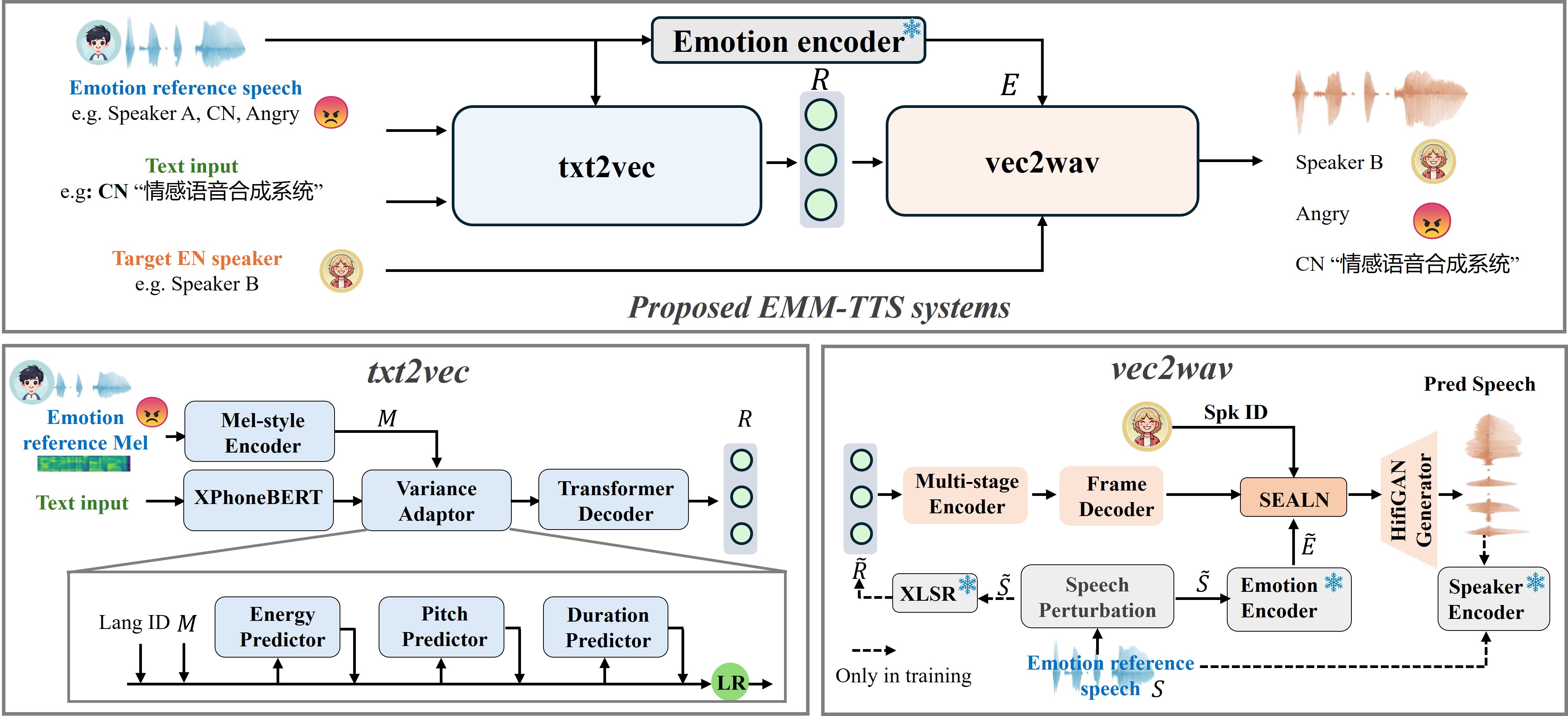}}
\centering
\caption{Overview of the proposed EMM-TTS systems. The top figure presents an overview of the entire framework. The lower-left part illustrates the emotion-dependent representation prediction module, while the lower-right part shows the speech generation module based on speaker-perturbation representations.}
\label{fig:2}
\end{figure*}
\subsection{Problem definition and model review}
Given a reference speech from speaker $B$ (e.g., in Chinese), our goal is to enable speaker $A$ (e.g., an English native speaker) to speak Chinese with the reference speech's emotion while retaining their own timbre, as depicted in the Figure \ref{fig:1}.
In contrast to the currently popular voice cloning methods \citep{10842513,chen2024f5,du2024cosyvoice,anastassiou2024seed}, which determine all attributes of the synthesized speech—such as emotion and timbre—based on a single reference audio, our research focuses on cross-lingual emotional speech synthesis. We enable independent control over both emotion and timbre.

A key challenge in cross-lingual TTS lies in decoupling speaker and emotion information. To address this, we propose a two-stage emotional speech synthesis system, EMM-TTS, shown in Figure \ref{fig:2}. The first stage \texttt{txt2vec} models and predicts emotions, while the second stage \texttt{vec2wav} controls speaker-specific characteristics.

\subsection{Emotion-dependent representations prediction}
As illustrated in Figure \ref{fig:2}, the \texttt{txt2vec} model proposed in this paper enhances the \texttt{txt2vec} model in ZMM-TTS \citep{10669054} and comprises an XPhoneBERT \citep{thenguyen23_interspeech} encoder, a Mel-style encoder, a variance adaptor, and a decoder for discrete SSL representations. 
In our study, the primary objective of \texttt{txt2vec} is to 
predict SSL representations $R$ with sufficient emotional information. To achieve this, the \texttt{txt2vec} model approaches emotion modeling from both implicit and explicit perspectives. 

For implicit information, we use a Mel-style encoder to learn a sentence-level global implicit style embedding $M$ that captures information such as speaker identity and emotion. The Mel-style encoder employs the same network architecture as described in \citep{pmlr-v139-min21b}, which comprises three main components: spectral processing, temporal processing, and multi-head self-attention. 

For explicit information, the values (pitch, energy, and duration) are extracted from paired text-speech data in training. And we use three predictors to infer the values. The pitch and energy predictors are both based on a two-layer 1D convolutional neural network using ReLU activation, followed by layer normalization, a dropout layer, and an additional linear layer as \cite{liu2021delightfultts}.
We employ a learnable aligner \citep{9747707} to estimate phoneme durations.
For language ID, another explicit information, we use a lookup embedding. 
It is important to note that modeling explicit information also relies on global implicit representations $M$ like Figure \ref{fig:2}.
\subsection{Speech generation via speaker-perturbation representations}
One of the fundamental challenges in Cross-lingual/speaker emotional TTS is the decoupling of timbre and style.
Considering that the representation $R$ predicted in the \texttt{txt2vec} stage contains sufficient emotional information, it also inevitably includes speaker information that is inconsistent with the target speaker's timbre. Therefore, we propose improvements to the \texttt{vec2wav} model in ZMM-TTS to address this issue, as shown in Figure \ref{fig:2}. First, we adopt speaker ID rather than pre-trained speaker representations, as we found that pre-trained representations inevitably lead to emotional information leakage. We then introduce a global emotional representation $E$ extracted from a pre-trained SSL-based emotion recognition model \citep{ma-etal-2024-emotion2vec}.

Specifically, in our approach, we perform a speaker perturbation, denoted as $sp()$, on the original waveform $Speech$ during training, which allows us to obtain a speaker-independent
signal denoted as $\widetilde{\text{Speech}}=sp(\text{Speech})$.
Subsequently, we extract the multilingual discrete SSL representation and emotion representation from the perturbed $\widetilde{\text{Speech}}$, denoted as $\widetilde{R}$ and $\widetilde{E}$. The perturbation processes for $R$ and $E$ are conducted independently. In this work, we explore two different speaker perturbation strategies. The first is signal-processing-based, implemented via formant shifting. The second uses speaker anonymization, generating speech with speaker characteristics that differ from those of the original audio. The process of formant shifting is illustrated in Algorithm \ref{alg:gender_shift}.

\begin{algorithm}[t]
\caption{Speaker Perturbation via Format Shift}
\label{alg:gender_shift}
\begin{algorithmic}[1]
\REQUIRE Source directory $\mathcal{D}_{\text{source}}$ with WAV files
\ENSURE Target directory $\mathcal{D}_{\text{target}}$ with manipulated WAV files

\FORALL{WAV file $f$ in $\mathcal{D}_{\text{source}}$}
    \STATE Load the sound signal $x$ from $f$
    \STATE Sample $s \sim \mathcal{U}(1, 1.4)$ and $s_1 \sim \mathcal{U}(0, 1)$
    \STATE $factor \leftarrow s$ if $s_1 \geq 0.5$, else $1/s$
    \STATE Extract pitch $p$ from $x$; compute median pitch $q$
    \STATE Manipulate $x' \leftarrow$ \textsc{ChangeFormant}$(x, factor)$
    \STATE Save $x'$ to $\mathcal{D}_{\text{target}}$
\ENDFOR
\end{algorithmic}
\end{algorithm}

Instead of adding or concatenating style embedding with encoder output, CLN \citep{9746858} and SALN \citep{pmlr-v139-min21b} use an element-wise product and a matrix addition. However, this approach only supports single-condition control. To address this limitation, we propose a multi-condition normalization mechanism that enables simultaneous control of emotion and timbre. 
This SEALN (Speaker-Emotion Adaptive Layer Normalization) takes the emotional representation \(E\) and the speaker representation \(S\) as inputs to predict the mean and standard deviation for the layer normalization of the frame decoder's output feature \(h\). 
Specifically, given a feature vector \(H = (h_1, h_2, \allowbreak \dots, h_D)\), where \(D\) is the dimensionality of the vector, the normalized feature vector \(Y = (y_1, y_2, \dots, y_D)\) is computed using the following equations: 
\begin{equation}
\begin{gathered}
\boldsymbol{y} = \frac{\boldsymbol{h} - \mu}{\sigma}, \quad \mu = \frac{1}{H} \sum^H_{i=1}\, h_i, \quad \sigma = \sqrt{\frac{1}{H}\sum^H_{i=1}(h_i-\mu)^2}
\end{gathered}
\end{equation}
Here, \(\mu\) and \(\sigma\) represent the mean and standard deviation of \(h\), respectively. Then, new 
\(\mu\) and \(\sigma\) values are computed based on the speaker representation \(S\) and emotional representation \(E\). The layers used to calculate the expected mean and standard deviation are simple fully connected layers, \(g()\) and \(b()\). Finally, the implementation process of SEALN is described as follows:
\begin{equation}
SEALN(\boldsymbol{h}, S, E) = g(S) \cdot \boldsymbol{y} + b(E)
\end{equation}
\(g(S)\) and \(b(E)\) adaptively scale and shift the normalized $h$ based on the speaker and emotional representation.
Using SEALN, it is possible to synthesize speech with varying emotions for different speakers under the given conditions of \(g(S)\) and \(b(E)\).

To ensure the \texttt{vec2wav} recovers the target timbre from the perturbed features and the speaker ID, we introduced a Speaker Consistency Loss (SCL), as described in paper \citep{pmlr-v162-casanova22a}.  
A pre-trained speaker encoder extracts speaker embeddings from the generated speech and the ground truth. We then maximize the cosine similarity as the speaker consistency loss. Let \( \phi(.) \) be a function that outputs the embedding of a speaker. Let \( cos\_sim \) denote the cosine similarity function, and let \( \alpha \) be a positive real number that controls the influence of the Speaker Contrastive Loss (SCL) in the final loss calculation. Additionally, let \( n \) represent the batch size. The SCL is defined as follows:
\begin{equation}
\label{eq:SCL}
     L_{SCL} =  \frac{- \alpha }{n} \cdot \sum_{i}^{n}{~cos\_sim(\phi(g_{i}), \phi(h_{i}))}
\end{equation}
where $g$ and $h$ represent, respectively, the ground truth and the generated speaker audio. 
Finally, the optimization objective of the entire \texttt{vec2wav} process consists of two components: reconstruction loss used in the original \texttt{vec2wav} of ZMM-TTS and speaker consistency loss.
\section{EXPERIMENTS}
\begin{table*}[t]
\centering
\caption{Details of the training corpora for the crosslingual model.}
\label{tab:1}
\begin{tabularx}{0.8\linewidth}{X|X|X|XXXXX}
\hline
\multirow{2}{*}{\textbf{Datasets}}&\multirow{2}{*}{\textbf{Language}}&\multirow{2}{*}{\textbf{Speaker}}&\multicolumn{5}{c}{\textbf{Emotions}}\\
\cmidrule(r){4-8}
& & &\textbf{{Neutral}} & \textbf{{Happy}} & \textbf{{Sad}} & \textbf{{Angry}} & \textbf{{Surprise}} \\
\hline
ESD\_ch &Chinese &10 &3,500 & 3,500 &3,500 &3,500 &3,500 \\
ESD\_en &English &10 &3,500 & 3,500 &3,500 &3,500 &3,500\\
Biaobei &Chinese &1 & 10,000 &0 &0 &0 &0 \\
LJSpeech &English &1  &13,100 &0 &0 &0 &0 \\
LibirTTS &English &1  &13,100 &0 &0 &0 &0 \\
\hline
\end{tabularx}
\end{table*}
\begin{table*}[t]
  \centering
  \caption{Voice cloning performance on LibriSpeech test-clean set.}
  \label{tab:voice_cloning}
  \setlength{\tabcolsep}{8pt}
  \begin{tabular}{lccccc}
    \toprule
   \textbf{Method} &\textbf{ WER (\%) $\downarrow$ } & \textbf{UTMOS $\uparrow$} & \textbf{SECS $\uparrow$} & \textbf{RTF$\downarrow$} & \textbf{Params$\downarrow$}\\
    \midrule
    HierSpeech++ \citep{11078430} &2.03  &4.40  & 0.591 &0.217 &204M \\
    ZMM-TTS \citep{10669054} & 2.37 &4.07  & 0.644 & 0.003 &167M  \\
    EMM-TTS & 2.28  &4.11 &0.661 &0.027 &183M\\
    \hline
   Ground-truth &2.14  &4.13  &- \\
    \bottomrule
  \end{tabular}
\end{table*}
This section describes the experimental data, preprocessing steps, and implementation details. 
The experimental data come from two languages—Chinese and English—and consist of publicly available datasets Biaobei\footnote{\url{https://www.data-baker.com/data/index/TNtts}}, LJSpeech \citep{ljspeech17}, LibriTTS \citep{zen19_interspeech}, and ESD \citep{ZHOU20221}.
We designed two categories of experiments: one to evaluate voice cloning performance in a monolingual setting, and the other to assess emotional speech synthesis in a cross-lingual scenario.

\subsection{Data and Preprocessing}
\textbf{Biaobei} dataset contains 10,000 utterances, totaling approximately 12 hours of Mandarin speech. The recordings were conducted in a professional studio using consistent equipment and software throughout the process, with a signal-to-noise ratio (SNR) of no less than 35 dB. The audio is recorded in mono at a sampling rate of 48 kHz, 16-bit resolution, and stored in PCM WAV format. It is one of the most widely used high-quality single-speaker datasets in speech synthesis.

\textbf{LJSpeech} is a publicly available speech dataset containing 13,100 short audio clips of a single speaker reading excerpts from seven non-fiction books. The clips range from 1 to 10 seconds in length and total approximately 24 hours.

\textbf{LibriTTS} consists of 585 hours of speech data at a 24kHz sampling rate from 2,456 speakers and the corresponding texts. The LibriTTS corpus is designed for TTS research.

\textbf{ESD} dataset contains 350 parallel utterances spoken by 10 native Mandarin speakers, and 10 English speakers with five emotional states (neutral, happy, angry, sad, and surprise).

For the voice cloning experiments in a monolingual setting, we used the LibriTTS dataset for training and the \textit{test-clean} subset of \textbf{LibriSpeech} \citep{7178964} for evaluation. This widely used test set contains speech from 40 different speakers and totals 5.4 hours of audio. Following the method described in \citep{ju2024naturalspeech}, we randomly evaluated 25 utterances per speaker from the LibriSpeech test-clean dataset.

For the cross-lingual emotional speech synthesis experiments, the ESD dataset has a limited size of 350 unique sentences per language. Therefore, training includes LJSpeech and Biaobei. To balance emotion and speaker representation, the ESD dataset is upsampled by a factor of 5 during training. The details of the training data in cross-lingual scenarios are shown in Table \ref{tab:1}.
\subsection{Model and Training Setup}
This subsection presents the details of two different experimental setups, including baseline models, evaluation metrics, and other relevant configurations.
\subsubsection{Monolingual Voice Coling}
This set of experiments primarily evaluates the model's performance in zero-shot speech synthesis. Accordingly, our proposed EMM-TTS uses a pretrained speaker embedding instead of a one-hot vector to represent speaker identity. The pretrained representation is the same as in \citep{10669054}, extracted from a pretrained ECAPA-TDNN model. Moreover, no information perturbation was applied to the data during training or inference.

\textbf{Reference Model.} 
For the monolingual voice cloning experiments, we compared our EMM-TTS against the following state-of-the-art (SOTA) models.
\begin{itemize}
    \item \textbf{HierSpeech++. \citep{11078430}} HierSpeech++ is a fast and efficient zero-shot speech synthesizer for text-to-speech that employs a hierarchical variational autoencoder. 
    Note that, for fair comparison, we did not use the super-resolution model. We used the official code and checkpoint for the experiments\footnote{\url{https://github.com/sh-lee-prml/HierSpeechpp}}.
    \item \textbf{ZMM-TTS.} ZMM-TTS is a multilingual, multispeaker framework with zero-shot generalization abilities for both unseen speakers and unseen languages.
\end{itemize}
While these models can synthesize multiple languages, we trained them solely on LibriTTS-960 to ensure fairness.
We chose LibriSpeech \citep{7178964} testclean as our benchmark dataset for the zero-shot TTS task.

\subsubsection{Cross-lingual Emotion Synthesis}
This set of experiments primarily evaluates the ability to transfer and synthesize emotions across languages. In these scenarios, our proposed EMM-TTS model adopts one-hot vectors as speaker input.
We experimented with two different speaker perturbation strategies. One based on signal processing, specifically formant perturbation, and the implementation of formant shifting followed the NANSY \citep{NEURIPS2021_87682805} model by using Praat \footnote{\url{https://www.fon.hum.uva.nl/praat/}}. 
The other uses an SSL-based language-independent speaker anonymization method by replacing the speaker embedding \citep{miao22_odyssey,10244064} and its official implementation\footnote{\url{https://github.com/nii-yamagishilab/SSL-SAS}}.
The proposed EMM-TTS model defaults to using formant shift as speaker perturbation, while an ablation experiment model, EMM-TTS-SA, is designed for speaker anonymization.

\textbf{Reference Model.} We refer to our proposed model as EMM-TTS and the two baseline models as DiCLET and M3:
\begin{itemize}
    \item DiCLET \citep{10244091}: This is a cross-lingual emotion transfer method based on a diffusion model that can transfer emotion from the source speaker to the target speaker, including both within-language and cross-lingual target speakers. Furthermore, to alleviate the entanglement among emotion, speaker, and language, multiple classification constraints, such as a speaker classifier and an emotion classifier, are employed, along with adversarial training.

    \item M3 \citep{shang2021incorporating}: M3 is a multi-speaker, multi-style, multilingual speech synthesis system based on FastSpeech, which incorporates a fine-grained style encoder to alleviate foreign accent issues.
    Emotion IDs and an emotion classifier are introduced into both the style predictor and style encoder to enable M3 for emotional transfer.
    
\end{itemize}

\subsection{Evaluation Metrics}
\begin{table*}[t]
\centering
\caption{Subjective evaluation results of Chinese speech (95\% confidence interval under t-distribution).}
\label{tab:3}
 \setlength{\tabcolsep}{8pt}
\begin{tabular}{lcccccc}
\toprule
\textbf{Model/Metric} & \multicolumn{3}{c}{\textbf{CN Speaker}} & \multicolumn{3}{c}{\textbf{EN Speaker}} \\
\cmidrule(r){2-4} \cmidrule(r){5-7}
 & \textbf{MOS} & \textbf{DMOS} & \textbf{EMOS} & \textbf{MOS} & \textbf{DMOS} & \textbf{EMOS} \\
\midrule
M3 \citep{shang2021incorporating} & 3.72$\pm$0.12 & 3.91$\pm$0.20 & 3.74$\pm$0.25 & 3.52$\pm$0.14 & 3.51$\pm$0.31 & 3.65$\pm$0.17 \\
DiCLET-TTS \citep{10244091} & 4.04$\pm$0.28 & 3.88$\pm$0.16 & 3.85$\pm$0.18 & 3.79$\pm$0.31 & 3.69$\pm$0.30 & 3.84$\pm$0.25 \\
EMM-TTS & 4.12$\pm$0.17 & 3.95$\pm$0.21 & 3.97$\pm$0.15 & 3.92$\pm$0.22 & 3.81$\pm$0.25 & 3.96$\pm$0.19 \\
GT & 4.63$\pm$0.13 &  &  &  &  &  \\
\bottomrule
\end{tabular}
\end{table*}
\begin{table*}[t]
\centering
\caption{Subjective evaluation results of English speech (95\% confidence interval under t-distribution).}
\label{tab:4}
 \setlength{\tabcolsep}{8pt}
\begin{tabular}{lcccccc}
\toprule
\textbf{Model/Metric} & \multicolumn{3}{c}{\textbf{CN Speaker}} & \multicolumn{3}{c}{\textbf{EN Speaker}} \\
\cmidrule(r){2-4} \cmidrule(r){5-7}
 & \textbf{MOS} & \textbf{DMOS }& \textbf{EMOS} & \textbf{MOS} & \textbf{DMOS} & \textbf{EMOS} \\
\midrule
M3 \citep{shang2021incorporating} & 3.42$\pm$0.14 & 2.98$\pm$0.11 & 3.01$\pm$0.37 & 3.64$\pm$0.17 & 3.78$\pm$0.13 & 3.67$\pm$0.15 \\
DiCLET-TTS \citep{10244091} & 3.67$\pm$0.18 & 3.59$\pm$0.12 & 3.62$\pm$0.22 & 3.81$\pm$0.31 & 3.90$\pm$0.26 & 3.73$\pm$0.20 \\
EMM-TTS  & 3.89$\pm$0.11 & 3.68$\pm$0.25 & 3.71$\pm$0.18 & 4.07$\pm$0.24 & 4.06$\pm$0.21 & 3.87$\pm$0.22 \\
GT & & & & 4.37$\pm$0.12 & & \\
\bottomrule
\end{tabular}
\end{table*}
\begin{table*}[t]
\centering
\caption{Objective evaluation results of Chinese and English speech synthesized by different systems.}
\label{tab:5}
 \setlength{\tabcolsep}{8pt}
\begin{tabular}{lcccccccc}
\toprule
& \multicolumn{4}{c}{\textbf{CN Speech}} & \multicolumn{4}{c}{\textbf{EN speech}}\\
\textbf{Model/Metric} & \multicolumn{2}{c}{\textbf{CN Speaker}} & \multicolumn{2}{c}{\textbf{EN Speaker}} & \multicolumn{2}{c}{\textbf{CN Speaker}} & \multicolumn{2}{c}{\textbf{EN Speaker}} \\
\cmidrule(r){2-3} \cmidrule(r){4-5} \cmidrule(r){6-7} \cmidrule(r){8-9}
 & \textbf{SECS} & \textbf{CER} & \textbf{SECS} & \textbf{CER}  & \textbf{SECS} &\textbf{ CER} & \textbf{SECS} & \textbf{CER} \\
\midrule
M3 \citep{shang2021incorporating} & 0.563 & 8.13 & 0.521 &10.03 &0.607 & 10.15 & 0.538 & 9.74 \\
DiCLET-TTS \citep{10244091} & 0.621 &9.92 & 0.557 &10.91 & 0.524 & 11.26 & 0.552 & 10.25 \\
EMM-TTS & 0.662 &7.13 & 0.643 & 7.47 & 0.597 & 8.90 & 0.614 & 8.21 \\
\bottomrule
\end{tabular}
\end{table*}
We analyzed the experimental results using both subjective and objective evaluations, with the following metrics included:

\paragraph{Objective evaluation} The objective metrics mainly evaluate the naturalness and similarity of the synthesized audio in both monolingual and cross-lingual experiments.
\begin{itemize}
    \item \textbf{SECS.} To assess speaker similarity, we compute SECS using the SOTA
speaker verification model, WavLM-Large \footnote{\url{https://github.com/microsoft/UniSpeech/tree/main/}}, to evaluate the speaker similarity, enabling comparison with those studies.
    \item \textbf{CER.} We employ whisper-large-v3 \footnote{\url{https://huggingface.co/openai/whisper-large-v3}} to transcribe the synthesized speech into text, which is then compared with the ground-truth transcripts to compute the character error rate (CER).
    \item \textbf{UTMOS.}  We adopt a state-of-the-art MOS prediction model, UTMOS \footnote{\url{https://github.com/sarulab-speech/UTMOS22}}, to objectively evaluate the naturalness of the generated audio.
    \item \textbf{EECS.} Similar to speaker similarity, we compute the emotional similarity of speech, where the emotion embeddings are extracted using the model emotion2vec \footnote{\url{https://github.com/ddlBoJack/emotion2vec}}.
\end{itemize}
In addition to evaluating speech quality, the proposed model’s complexity is assessed based on the real-time factor \textbf{(RTF)} and the number of parameters \textbf{(Params)}. RTF measures the time required to generate one second of audio on a GPU. In this experiment, RTF is tested on a single NVIDIA RTX 4090 GPU with 24 GB of memory.

\paragraph{Subjective evaluation}
Considering that objective metrics in cross-lingual scenarios may fail to capture subtle variations in emotion and speaker characteristics, we further conducted the following subjective experiments.
\begin{itemize}
    \item \textbf{MOS.} The Mean Opinion Score (MOS) is employed to evaluate the naturalness of audio, ranging from 1 to 5, where 1 indicates very poor quality and 5 indicates excellent quality.
    \item \textbf{DMOS.} The Differential Mean Opinion Score (DMOS) is employed to evaluate the speaker similarity between synthesized and reference audio, on a 1–5 scale where 1 denotes completely dissimilar and 5 denotes highly similar.
    \item \textbf{EMOS.} The Emotion Mean Opinion Score (EMOS) is employed to evaluate the emotional similarity between synthesized and reference audio, on a 1–5 scale where 1 denotes completely dissimilar and 5 denotes highly similar.
    \item \textbf{ABX test.} The ABX test is employed to evaluate perceptual preference by asking listeners to judge which of two audio samples exhibits higher naturalness or greater similarity. Listeners may also indicate that the two samples are indistinguishable.
\end{itemize}
In the subjective evaluation, each system generates 30 sentences for each language. These include six speakers, each contributing one sentence for each of five emotions. A total of 15 participants were invited to evaluate the subjective tests.

\section{Experiment Results}
In this section, we validate the effectiveness of the proposed method in both monolingual and cross-lingual scenarios. In the monolingual setting, we primarily analyze the performance of voice cloning; in the cross-lingual setting, we also evaluate emotional similarity. We further investigate the impact of speaker perturbations on the model and conduct ablation studies on SEALN and SCL.

\subsection{Performance on monolingual voice cloning}
The results of EMM-TTS and the baseline models on the LibriSpeech test-clean set are presented in Table \ref{tab:voice_cloning}. 
Compared with ZMM-TTS, incorporating both explicit and implicit emotional representations enables EMM-TTS to achieve higher speaker similarity and improved speech naturalness.
This suggests that, in addition to timbre cloning, modeling emotional information can substantially improve speaker similarity with the reference audio.
Compared with the current state-of-the-art multilingual synthesis model HierSpeech++, EMM-TTS achieves a notable improvement in speaker similarity under the same training data conditions.
By comparing the RTF and the number of parameters with HierSpeech++, our model is more lightweight and better suited for computation-constrained environments.

\subsection{Performance on cross-lingual emotion speech synthesis}
\subsubsection{Compare with baseline methods}
\paragraph{Subjective results}
Tables \ref{tab:3} and \ref{tab:4} present the subjective evaluation results of synthesized Chinese and English speech, respectively. The proposed EMM-TTS achieves the best naturalness. This improvement may be attributed to the XPhoneBERT-based text representation and phoneme encoder, which enable more effective modeling of pronunciations from different languages in a unified space, thereby enhancing multilingual synthesis capability. Furthermore, we find that the naturalness degrades significantly when synthesizing speech with cross-lingual speakers compared to same-lingual speakers. Specifically, when synthesizing text in language B with the voice of a speaker from language A, the generated speech often contains pronunciation errors and accent issues, particularly when English speakers synthesize Chinese speech.

For \textbf{DMOS}, DiCLET-TTS and M3 achieve relatively similar results under monolingual conditions, but M3 exhibits a substantial performance drop in cross-lingual scenarios. This indicates that M3 suffers from weak disentanglement capability. When the reference audio and the target speaker’s timbre are mismatched, the synthesized speech is heavily affected by the timbre of the reference audio. In contrast, the proposed \textbf{EMM-TTS} consistently achieves the best speaker similarity and emotion similarity in both same-lingual and cross-lingual settings, while also showing the least performance degradation in cross-lingual scenarios. These results demonstrate the effectiveness of our proposed emotion modeling and disentanglement strategies.

\paragraph{Objective results.} From the objective metrics reported in Table \ref{tab:5}, we observe that EMM-TTS achieves the best performance in both intelligibility and SECS across the two languages. Moreover, consistent with the subjective evaluations, when the target speaker’s language differs from the synthesized speech, both speaker similarity and intelligibility decline. In contrast, DiCLET-TTS consistently yields the poorest intelligibility (CER) in most cases, which may be attributed to its use of speaker-adversarial learning for text representations, potentially compromising the content quality of the synthesized speech.
\begin{figure}[p]
  \centering
  \begin{subfigure}{0.48\textwidth}
    \centering
    \includegraphics[width=\linewidth]{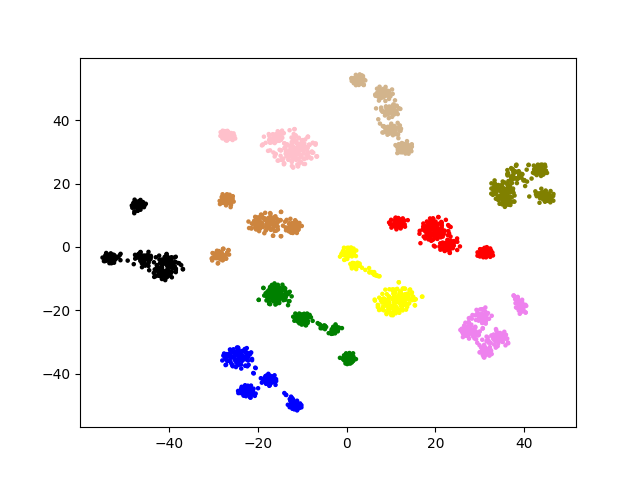}
    \caption{Original audio.}
    \label{fig:top}
  \end{subfigure}
  \hfill
  \begin{subfigure}{0.48\textwidth}
    \centering
    \includegraphics[width=\linewidth]{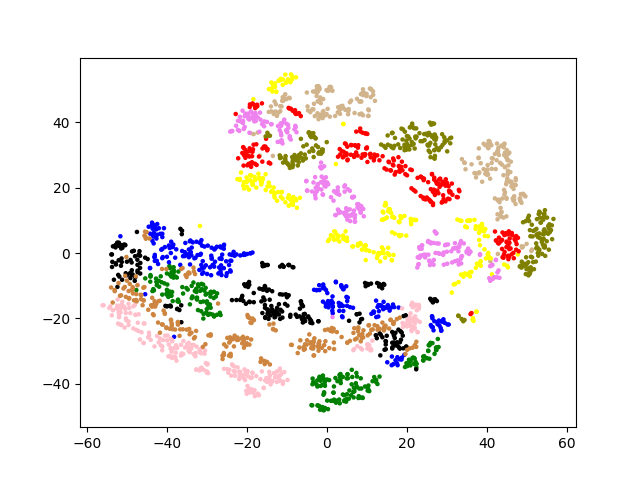}
    \caption{Formant shift.}
    \label{fig:middle}
  \end{subfigure}
  \hfill
  \begin{subfigure}{0.48\textwidth}
    \centering
    \includegraphics[width=\linewidth]{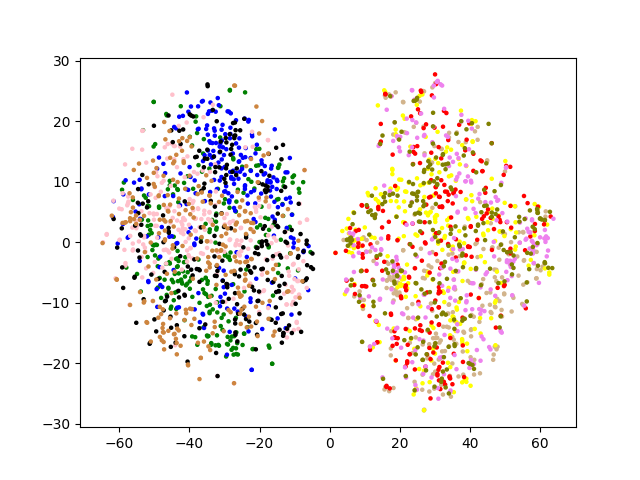}
    \caption{Speaker anonymization.}
    \label{fig:bottom}
  \end{subfigure}

  \caption{Visualization of speaker embeddings under different speaker perturbation conditions. Different colors representing different speak
ers.}
  \label{fig:3}
\end{figure}

\subsubsection{Analysis of the effectiveness of speaker perturbation}

\begin{figure}[ht]
\centerline{\includegraphics[width=\columnwidth]{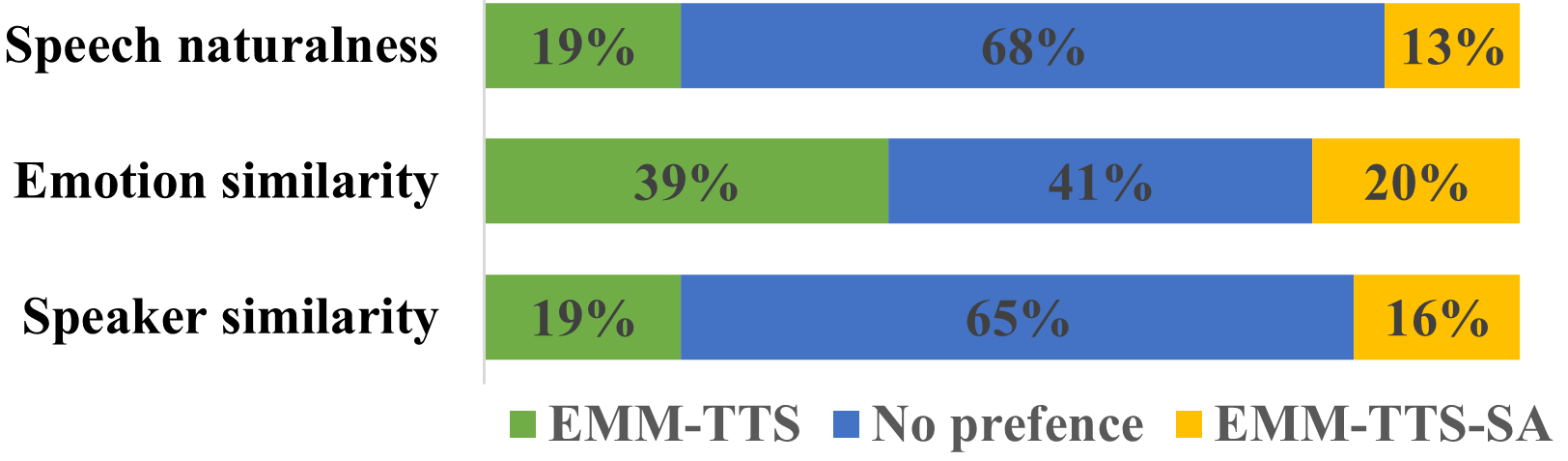}}
\caption{ABX test results for speech synthesized by the EMM-TTS model using two different speaker perturbation strategies.}
\label{fig:4}
\end{figure}

In addition to formant shifting, this chapter utilizes a speaker anonymization technique to alter information, aiming to investigate the effects of various interference methods on synthesized speech. First, audio samples from 10 Chinese speakers in the ESD dataset were selected for two types of speaker interference, followed by visualization and quantitative analysis of the interfered audio. 
For each speaker and each emotion, 50 sentences were selected, resulting in a total of 2,500 sentences for analysis.
Figure \ref{fig:3} presents a visualization of the speaker representations extracted by a pre-trained ECAPA-TDNN speaker encoder. The representations were reduced to two dimensions using t-SNE, with different colors representing different speakers. From Figure \ref{fig:3} (a), it can be observed that, in the original audio, speaker embeddings of the same speaker cluster closely together, forming distinct clusters. Furthermore, each speaker cluster contains several sub-clusters, which, upon inspection, correspond to different emotions. This phenomenon further confirms that speaker information and emotional information are often entangled. Although ECAPA-TDNN achieves good performance in speaker classification, its learned speaker representations still contain rich emotional details. Furthermore, as shown in Figures \ref{fig:3} (b) and \ref{fig:3} (c), speaker interference methods can effectively alter the speaker information in the audio. Specifically, after interference, the embeddings of audio samples from the same speaker exhibit greater distances. Among the two methods, speaker anonymization imposes the greatest interference with speaker information.

In addition to the visual analysis, Table \ref{tab:6} presents the objective evaluation results for audio processed with different speaker perturbation methods. The SECS results are consistent with the observations in Figure \ref{fig:3}, showing that both perturbation methods effectively interfere with speaker-related information in the audio. The anonymization-based method produces the most substantial perturbation to speaker identity, but it also inevitably degrades emotional expressiveness. This method, while most effective at obfuscating speaker identity, introduces the greatest emotional distortion.
Analysis of the UTMOS and CER values further reveals that the formant-shift method primarily affects the naturalness of speech, whereas the anonymization-based method mainly impacts the linguistic content. On one hand, directly shifting formants tends to make the speech sound less natural. On the other hand, the anonymization approach relies on recognizing and re-synthesizing the speech, and recognition errors can easily accumulate in the anonymized output.

\begin{table}[t]
\centering
\caption{Objective evaluation of speech after applying two different speaker perturbations.}
\label{tab:6}
 \setlength{\tabcolsep}{3pt}
\begin{tabular}{lllll}
\toprule
\textbf{Method} & \textbf{SECS} & \textbf{EECS} & \textbf{CER (\%)} & \textbf{UTMOS} \\
\midrule
Formant shift & 0.514  & \textbf{0.848}  &\textbf{9.07}    &2.163 \\
Speaker anonymization & \textbf{0.354}  & 0.799  &20.57   &\textbf{3.055} \\
Original audio  &1.000   & 1.000  &4.88    &2.907 \\
\bottomrule
\end{tabular}
\end{table}

\begin{table*}[t]
\centering
\caption{Objective metrics of synthesized speech under different speaker perturbation methods.}
\label{tab:7}
\setlength\tabcolsep{8pt}
\begin{tabular}{lcccccccc}
\toprule
& \multicolumn{4}{c}{\textbf{CN Speech}} & \multicolumn{4}{c}{\textbf{EN Speech}}\\
\textbf{Model/Metric} & \multicolumn{2}{c}{\textbf{CN Speaker}} & \multicolumn{2}{c}{\textbf{EN Speaker}} & \multicolumn{2}{c}{\textbf{CN Speaker}} & \multicolumn{2}{c}{\textbf{EN Speaker}} \\
\cmidrule(r){2-3} \cmidrule(r){4-5} \cmidrule(r){6-7} \cmidrule(r){8-9}
 & \textbf{SECS }& \textbf{CER} & \textbf{SECS} & \textbf{CER}  & \textbf{SECS} &\textbf{ CER} & \textbf{SECS} & \textbf{CER }\\
\midrule
EMM-TTS (w/ formant shift)    & 0.662 & 7.13 & 0.643 & 7.47 & 0.597 & 8.90 & 0.614 & 8.21 \\
EMM-TTS (w/ speaker anonymization)  & 0.657 & 12.24 & 0.541 & 11.13 & 0.550 & 11.30 & 0.617 & 10.37 \\
EMM-TTS (w/o speaker erturbation) & 0.627 & 7.08 & 0.503 & 7.44 & 0.532 & 9.12 & 0.603 & 8.07 \\
\bottomrule
\end{tabular}
\end{table*}

Table \ref{tab:7} presents the objective evaluation results of the EMM-TTS model under different speaker perturbation strategies. Compared with the model that applies no speaker perturbation, introducing perturbations reduces the reference speaker's influence on the synthesized audio, leading to improved SECS. This result indicates that perturbing speaker information facilitates disentangling emotion from speaker identity. Although the two perturbation methods yield comparable SECS scores, the anony\allowbreak{}mization-based perturbation causes a noticeable decline in speech intelligibility.

Figure \ref{fig:4} illustrates the ABX test preferences for the EMM-TTS (default with formant shift) model when different speaker perturbation methods are applied. The test was conducted on samples spoken by English speakers with Chinese linguistic content. The results show that the formant-shift method outperforms the anonymization-based approach in terms of naturalness, speaker similarity, and emotional similarity. Among these aspects, the gap in emotional similarity is the most pronounced. Although the anonymization-based method effectively disrupts speaker identity, it also weakens emotional cues, leading to synthesized speech that sounds more neutral.

\begin{table*}[t]
\centering
\caption{Subjective evaluation results of synthesized Chinese speech across different models.}
\label{tab:8}
 \setlength{\tabcolsep}{12.7pt}
\begin{tabular}{lcccccc}
\toprule
\textbf{Model/Metric} & \multicolumn{3}{c}{\textbf{CN Speaker}} & \multicolumn{3}{c}{\textbf{EN Speaker}} \\
\cmidrule(r){2-4} \cmidrule(r){5-7}
 & \textbf{MOS }&\textbf{ DMOS} & \textbf{EMOS} &\textbf{ MOS} & \textbf{DMOS} & \textbf{EMOS} \\
\midrule
EMM-TTS &4.12$\pm$0.17 & 3.95$\pm$0.21 & 3.97$\pm$0.15 & 3.92$\pm$0.22 & 3.81$\pm$0.25 & 3.96$\pm$0.19 \\
 \quad w/o SCL   & 4.09$\pm$0.22 & 3.87$\pm$0.21 & 4.02$\pm$0.23 & 3.89$\pm$0.19 & 3.58$\pm$0.15 & 4.10$\pm$0.27 \\
\quad w/o emo   & 4.13$\pm$0.21 & 4.09$\pm$0.13 & 3.86$\pm$0.20 & 3.90$\pm$0.14 & 3.88$\pm$0.22 & 3.87$\pm$0.13 \\
 \quad w/o SSALN & 4.10$\pm$0.23 & 3.82$\pm$0.30 & 4.02$\pm$0.13 & 3.99$\pm$0.18 & 3.61$\pm$0.11 & 3.97$\pm$0.24 \\
\bottomrule
\end{tabular}
\end{table*}

\subsubsection{Ablation Study}
In the proposed vec2wav model, several additional modules are incorporated, including the Speaker Consistency Loss (SCL), the Speaker-Emotion Adaptive Layer Normalization \allowbreak{}(SEALN), and the pretrained emotional representation 
E. The subjective ablation results of these modules are presented in Table \ref{tab:8}. As shown in the table, both the SCL constraint and the SSALN module play a crucial role in maintaining similarity to the target speaker. Although speaker perturbation is applied during training, these components enable the model to recover accurate speaker identity from the perturbed representations.

Moreover, removing the pretrained emotional representation 
E leads to a noticeable decrease in emotional similarity. Interestingly, emotional similarity and speaker similarity tend to exhibit a negative correlation—improving one often comes at the cost of the other. The final EMM-TTS model achieves a balanced trade-off between the two, demonstrating superior overall performance. Future work will explore finer-grained control over both timbre and emotional expressiveness, aiming to achieve a more flexible balance between them.

\section{Discussion}

In this work, we propose a two-stage cross-lingual emotional speech synthesis system, EMM-TTS. The first stage focuses on modeling and predicting emotional representations, while the second stage enables fine-grained control over speaker timbre. The two stages are connected through perturbed self-supervised features, which serve as a bridge between emotion and timbre modeling. Experimental results demonstrate that EMM-TTS achieves strong zero-shot voice cloning in monolingual scenarios and effective emotion transfer across languages.

Timbre and emotion are two highly entangled factors in speech signals, posing challenges for fine-grained control in speech synthesis. Information perturbation is a commonly adopted strategy for disentangling these factors. Previous studies have primarily focused on perturbation methods based on signal processing. In this work, we investigate the capability of recent speaker anonymization models to disentangle emotion and timbre. Our analysis combines visualization, subjective listening tests, and objective audio quality metrics. Experimental results show that signal-processing-based perturbations produce stronger distortion of speaker identity, whereas speaker anonymization models better preserve the naturalness of synthesized speech.

Our study further reveals that pretrained features—such as high-dimensional latent variables learned by speaker or emotion encoders—cannot fully replace explicit acoustic features such as pitch, energy, and duration. Experimental results show that incorporating the modeling and prediction of these explicit features enhances the model’s voice cloning capability. In the emotion transfer stage, we introduce the Speaker Consistency Loss (SCL) and the Speaker-Emotion Adaptive Layer Normalization (SEALN). The ablation results demonstrate that these components contribute positively to maintaining speaker timbre and improving the overall synthesis quality.

\section{Conclusion}
In this work, we proposed EMM-TTS, a two-stage cross-lingual emotional text-to-speech system that effectively disentangles emotion and timbre through speaker-perturbed SSL representations. By leveraging explicit prosodic modeling in the first stage and timbre restoration in the second stage, the system enables controllable emotion transfer and high-fidelity speaker imitation across languages. The proposed Speaker Consistency Loss (SCL) and Speaker–Emotion Adaptive Layer Normalization (SEALN) further enhance timbre stability and expressive consistency. Moreover, experiments reveal that combining explicit acoustic features with pretrained latent representations improves timbre reproduction. Extensive subjective and objective evaluations confirm that EMM-TTS achieves superior performance in both zero-shot timbre cloning and cross-lingual emotion transfer. In future work, we will explore finer-grained control of emotion intensity and timbre style, as well as adaptive balancing strategies between emotional expressiveness and speaker identity.

Future work will explore more fine-grained control over both emotion and timbre, enabling continuous adjustment of emotional intensity and timbre style. We also plan to investigate adaptive mechanisms that can balance the trade-off between emotional expressiveness and speaker identity preservation. Extending the approach to support more languages and diverse emotional expressions will further enhance the generalization and applicability of the proposed EMM-TTS framework.

\section{Acknowledgments}
This work was supported in part by the National Natural Science Foundation of China under Grant (U23B2053, 62176182).

\bibliography{ref}

\end{document}